# Consistent Anomalies in Translation-Invariant Noncommutative Gauge Theories

Amir Abbass Varshovi

amirabbassv@ipm.ir

*School of Physics, Institute for Research in Fundamental Science (IPM)*

*P. O. Box 19395-5531, Tehran, IRAN*

*Department of Physics, Sharif University of Technology*

*P. O. Box 11365-9161, Tehran, IRAN*

**Abstract;** Translation-invariant noncommutative gauge theories are discussed in the setting of matrix modeled gauge theories. Using the matrix model formulation the explicit form of consistent anomalies and consistent Schwinger terms for translation-invariant noncommutative gauge theories are derived.

## 1. Introduction

Studying quantum field theories over noncommutative spaces has been intensively worked out in the last few decades. It has been shown that noncommutative Yang-Mills (NCYM) theories, even noncommutative QED (NCQED), share many properties of standard commutative non-Abelian gauge theories [1-5] provided the loop calculations in perturbation theory are referred to noncommutative vertices with momenta dependent phase factors which lead to similarities with ordinary non-Abelian gauge theories.

Actually, appearance of these nontrivial phase factors due to noncommutative products in the Feynman diagram vertices changes the well-defined behaviors of gauge theories in short and long distances which affect the causality [6-9], the locality [8-12], the unitarity [13-18], and eventually the renormalizability [19-31] of gauge theories. In fact, appearance of the momenta dependent phase factors in noncommutative vertices breaks down the loop diagrams into two substantially different forms; a) the diagrams with loop momenta independent phase factors which are called planar diagrams and b) the diagrams with loop momenta dependent phase factors, usually referred to as non-planar diagrams [32]. The appearance of non-planar diagrams causes the unfamiliar pathological behavior of UV/IR-mixing which threatens the renormalizability of noncommutative gauge theories. Actually it is known that in addition to NCYM theories, noncommutative scalar field theories also suffer from UV/IR mixing in renormalizability [19]. Although, UV/IR mixing in noncomutative $\phi^4$ theory can be cured in renormalization programs but it is known that this costs missing either the translation-invariance and consequently the energy-momentum conservation law [27, 28] or the locality of the theory [31]. Therefore, noncommutative field theories can be considered as one of the most challenging topics in theoretical high energy physics.

Indeed, UV/IR mixing is the phenomenon of appearance of IR divergences in loop calculations of Feynman diagrams with UV singularities. This phenomenon leads one to use simultaneously UV and IR regularizing methods to control the singularities of non-planar diagrams. Specially, the crucial point is that there exists a certain duality between ultraviolet and infrared behavior of non-planar diagrams in perturbative NCYM theories. This duality manifests itself in the singularity of amplitudes in two limits of small noncommutativity parameter $\theta$ of Groenewold-Moyal star product and large momentum cutoff $\Lambda$ of the theory [19].

The anomalous behaviors and eventually the renormalizability of NCYM theories have been extensively studied since the ending years of 1990s [33-45] but it is still an open problem in the realm of translation-invariant noncommutative gauge (TNG) theories [46, 47]. Despite of the vast amount of works in the last decade, an explicit description of chiral anomaly and Schwinger term in TNG theories, even Groenewold-Moyal QED has not been clearly found yet [33-45]. Actually, for a general translation-invariant star product, there are four different currents in NCQED with each one obeying a special classical conservation law [35, 36]. These four currents are gauge invariant and gauge covariant currents and their axial counter parts. Indeed, noncommutativity of gauge symmetry leads to two different kinds of Noether's currents which are either invariant or covariant under noncommutative gauge transformations. The most challenging problem is explicitly finding the anomalous behaviors of; a) the axial invariant current $j^{5\mu} := \bar{\psi}_\alpha \star \psi_\beta (\gamma^\mu \gamma^5)_{\alpha\beta}$ which is classically conserved, $\partial_\mu j^{5\mu} = 0$, and; b) the axial covariant current $J^{5\mu} := \psi_\beta \star \bar{\psi}_\alpha (\gamma^\mu \gamma^5)_{\alpha\beta}$. The classical conservation law for covariant current is similar to that of gauge current in non-Abelian Yang-Mills theories;

$$D(J^5) := \partial_\mu J^{5\mu} + i[J^{5\mu}, A_\mu]_\star = \partial_\mu J^{5\mu} + i(J^{5\mu} \star A_\mu - A_\mu \star J^{5\mu}) = 0$$

(1)

It has been shown [35, 36, 39, 41, 44] that the anomalous behavior of axial invariant current survives only on the zero mode of the Fourier expansion. Actually axial invariant current is anomalous only over a Lebesgue null set of the phase space. Therefore, it was believed that $j^5$ is anomaly free [39];

$$\partial_\mu \langle j^{5\mu} \rangle = 0$$

(2)

On the other hand, the familiar form of chiral anomaly with the ordinary product replaced by Groenewold-Moyal star product was discovered [35-37] for the anomalous behavior of axial covariant current;

$$D(\langle J^5 \rangle) = \frac{1}{16\pi^2} \epsilon^{\mu\nu\sigma\lambda} F_{\mu\nu} \star F_{\sigma\lambda}$$

(3)

for $F_{\mu\nu} = \partial_\mu A_\nu - \partial_\nu A_\mu + [A_\mu, A_\nu]_\star$ the noncommutative strength field. On the other hand it was shown [44] that considering the commutativity of time (i.e. $\theta^{0i} = 0$ for $i = 1,2,3$), (2) and (3) lead to a contradiction. In order to solve the paradox, the compactification of space was proposed in [45] as a remedy. Essentially the compactification of space converts the concrete phase space to a discrete lattice due to Fourier modes of periodic functions.

Therefore, integration over the phase space is replaced by a discrete summation which never kills the functions which live only on a single mode. The point of which solved the paradox of [44].

In this article we follow the method of matrix modeled gauge (MMG) theories [48] based on differential graded algebras and quantum groups to study anomalous behaviors of TNG theories as the asymptotic formulation of the first approximation of MMG theories. The idea of modified partition functions [48, 49] for MMG theories enables us to explicitly derive the anomalies of invariant and covariant currents for any translation-invariant star product. Following [50] the anomalous behaviors of axial currents can be precisely worked out from those of vector currents. Specially, using the matrix model formulation, it is shown that the invariant and covariant currents are two different expressions of a single current. Therefore, it is natural that they must have the same anomalous behaviors, the point of which illustrates the contradiction of [44] for any NCQED with a general translation-invariant star product. In fact, the anomalous behaviors of invariant and covariant currents for any translation-invariant NCQED are derived precisely and it is shown that they are completely similar to those of Groenewold-Moyal one.

Finally, the Schwinger term of the commutator of invariant and covariant currents for a general translation-invariant star product is worked out in the setting of MMG theories and the agreement of the result with that of loop calculations for Groenewold-Moyal NCQED [45] is discussed.

In section 2, a short summery of matrix model formulation of gauge theories is given [48]. In section 3, the anomalies of invariant and covariant currents and the Schwinger term of commutator of invariant and covariant currents are explicitly derived in the setting of MMG theories.

## 2. Matrix Modeling of Translation-Invariant Noncommutative Gauge Theories

In this section TNG theories are studied in the setting of matrix model formulation [48] over space-time $M = \mathbb{R}^{d_1} \times \mathbb{T}^{d_2}$, $d_1 \geq 1$, with noncompact time coordinate. Initially it should be reminded that the star product $\star$ on $C^\infty(\mathbb{R}^m)$, is translation-invariant if;

$$\mathcal{T}_a(f) \star \mathcal{T}_a(g) = \mathcal{T}_a(f \star g),$$

(4)

for any vector $a \in \mathbb{R}^m$ and for any $f, g \in C^\infty(\mathbb{R}^m)$, where $\mathcal{T}_a$, is the translating operator; $\mathcal{T}_a(f)(x) = f(x + a)$, $f \in C^\infty(\mathbb{R}^m)$. Replacing $a$ with $ta$, $t \in \mathbb{R}$, in (4) and differentiating with respect to $t$ at $t = 0$, one easily finds that $\partial_\mu(f \star g) = \partial_\mu f \star g + f \star \partial_\mu g$, which shows that any translation-invariant product is exact [48]. To be precautious and to have well-defined products, from now on $C^\infty(\mathbb{R}^m)$ is replaced by $\mathcal{S}(\mathbb{R}^m)$ for any translation-invariant product $\star$.

An equivalent definition of translation-invariant products over Cartesian space $\mathbb{R}^m$, is given by [46];

$$(f \star g)(x) := \int \frac{d^m p}{(2\pi)^m} \frac{d^m q}{(2\pi)^m} \tilde{f}(q) \tilde{g}(p) e^{\alpha(p+q,p)} e^{i(p+q).x},$$

(5)

for $f, g \in \mathcal{S}(\mathbb{R}^m)$, their Fourier transformations $\tilde{f}, \tilde{g} \in \mathcal{S}(\mathbb{R}^m)$, and finally for a 2-cycle $\alpha \in C^\infty(\mathbb{R}^m \times \mathbb{R}^m)$ which obeys the following cyclic property;

$$\alpha(p, r + s) + \alpha(r + s, r) = \alpha(p, r) + \alpha(p - r, s),$$

(6)

for any $p, r, s \in \mathbb{R}^m$. Actually (6) is equivalent to associativity of $\star$ and defines a cohomology theory to classify the translation-invariant products up to commutativity [46, 48]. In fact, (6) is basically reflected by condition $(f \star g) \star h = f \star (g \star h)$ for any three Schwartz functions $f, g$ and $h$. On the other hand, it can be easily checked that the definition (5) defines a translation-invariant product in agreement with definition (4). Thus, (5) is the most general form of translation-invariant deformation quantization of $C^\infty(\mathbb{R}^m)$. In the following we are looking for translation-invariant products on $C^\infty(M)$ which lead to nontrivial products on $C^\infty(\mathbb{T}^{d_2})$. More precisely it is supposed that the compactified coordinates are the responsible of noncommutativity of the space-time $M$. To study the noncommutative structures on $\mathbb{T}^m$ it is enough to restrict ourselves to periodic functions of $C^\infty(\mathbb{R}^m)$. Therefore, for translation-invariant products on $C^\infty(\mathbb{T}^m)$, the integrations in (5) should be replaced by discrete summations on the lattice of Fourier modes on $\mathbb{T}^m$.

Studying noncommutative gauge theories with translation-invariant star products in the setting of matrix model formulations can produce a collection of spectacular results. One of the standard definitions of TNG theories over $\mathbb{R}^{2n}$ is given by the following Lagrangian density;

$$\mathcal{L}_{Matter}(\psi) = \bar{\psi}_{\alpha,i} \star i\partial_\mu \psi_{\beta,i} \gamma^\mu_{\alpha\beta} - T^a_{ij}\, \gamma^\mu_{\alpha\beta}\, \bar{\psi}_{\alpha,i} \star A^a_\mu \star \psi_{\beta,j},$$

(7)

with $\star$ a translation-invariant product over $C^\infty(\mathbb{R}^{2n})$ and $T^a$s the Hermitian matrices as the represented generators of the Lie algebra. Obviously, for NCQED the color matrices $T^a$, disappear in (7). On the other hand, the noncommutativity of $\star$ gives rise to a number of different definitions for TNG theories. For example a natural alternative form for definition (7) is;

$$\mathcal{L}_{Matter}(\psi) = \bar{\psi}_{\alpha,i} \star i\partial_\mu \psi_{\beta,i} \gamma^\mu_{\alpha\beta} + T^a_{ij}\, \gamma^\mu_{\alpha\beta}\, \bar{\psi}_{\alpha,i} \star \psi_{\beta,j} \star A^a_\mu.$$

(8)

The expression (8) of NCYM theories, originally contains the invariant currents, while the covariant currents only appear in the action of (7) due to the trace property of $\star$. On the other hand, NCYM theories can be also defined in another form;

$$\mathcal{L}_{int}(\psi) = -T^a_{ij}\, \gamma^\mu_{\alpha\beta}\, \bar{\psi}_{\alpha,i} \star [A^a_\mu, \psi_{\beta,j}]_\star.$$

(9)

It is known [48] that the Lagrangian densities of (7)-(9) can all be unified in the formalism of matrix modeled gauge theories over $M$. Actually, it is possible for any translation-invariant star product, $\star$, on the set of smooth functions on space-time manifold $M = \mathbb{R}^{d_1} \times \mathbb{T}^{d_2}$ in a concrete taking limit procedure. More precisely, there exist a family of $\mathbb{C}$-linear maps, $\{\mathfrak{m}_N\}_{N\in\mathbb{N}}$, from $C^\infty(M)$ to $C^\infty(M) \otimes M_N(\mathbb{C})$ which each of $\mathfrak{m}_N$s leads to a truncated form of star product $\star$, denoted by $\star_N$, $N \in \mathbb{N}$, due to

$$\xi_N tr\{\mathfrak{m}_N(f_1)\dots\mathfrak{m}_N(f_n)\} = \int_{\mathbb{T}^{d_2}} f_1 \star_N \dots \star_N f_n,$$
(10)

for some model dependent factor $\xi_N$, such that $\star_N$ tends to $\star$ uniformly as $N \to \infty$. To formulate this theorem, a list of notations should be set;

- a) Let $(\mu) = (\mu_1, \dots, \mu_m)$ be an $m$-plet of integers. Then, set $|\mu| = m$, the length of $(\mu)$. For the empty multi-plet $(\emptyset) = (\ )$, set $|\emptyset| = 0$.
- b) Assume that $k \in \mathbb{R}^l$, $l \in \mathbb{N}$, is an arbitrary vector and $(\mu)$ is a multi-plet of coordinates with $|\mu| = m$. Then, by convention, $k^{(\mu)}$ means $k^{\mu_1} \times \dots \times k^{\mu_m}$. One also needs $k^{(\emptyset)}$ to be equal to unity for any $k$. Moreover, one can naturally generalize this notation for any arbitrary tensor with finitely many up and down indices. For example if $\Lambda$ is a matrix with up indices and $(\mu)$ and $(\nu)$ are two arbitrary multi-plets, then;

$$\Lambda^{(\mu)(\nu)} = \begin{cases} \Lambda^{\mu_1\nu_1} \times \dots \times \Lambda^{\mu_m\nu_m} & \text{if } |\mu| = |\nu| = m \\ 0 & \text{otherwise} \end{cases}.$$
(11)

  Also one demands for $\Lambda^{(\emptyset)(\emptyset)} = 1$ similarly.
- c) As mentioned above the space-time manifold is considered to be the Cartesian product space $\mathbb{R}^{d_1} \times \mathbb{T}^{d_2}$. It is also supposed that $d_2 = 2d$ and $d_1 + d_2 = 2n$. Indeed, $M = \mathbb{R}^{2(n-d)} \times \mathbb{T}^{2d}$ is the underlying space-time manifold. Moreover, it is also supposed that the radius $R$ of $\mathbb{T}^{2d}$ tends to infinity; $R \to \infty$. As it is seen in the following, for the case of Groenewold-Moyal noncommutative fields, $\mathbb{T}^{2d}$ plays the role of noncommutative torus $\mathbb{T}^{2d}_\theta$ for TNG theories in large $N$ limit of matrix models.
- d) Consider $\alpha$ from (5). Set $\alpha(p+q,p) = \langle \alpha_L(ip)|\alpha_R(iq)\rangle$, $p,q \in \mathbb{R}^{2n}$, for two either finite or possibly infinite dimensional vector valued functions $|\alpha_L\rangle$ and $|\alpha_R\rangle$. Actually the Taylor expansion formula for $\alpha$ with radius of convergence $\infty$, produces such vector valued functions. Thus, for infinite dimensional vectors $|\alpha_L\rangle$ and $|\alpha_R\rangle$, it should be supposed that the entries $\alpha_L^k$ and $\alpha_R^k$, tend to zero rapidly as $k \to \infty$. Here by rapidly decaying we mean; $\lim_{k\to\infty} k^s \alpha_L^k(ip)\alpha_R^k(iq) = 0$ for any $p,q \in \mathbb{R}^{2n}$ and for any $s \in \mathbb{N}$. On the other hand, it can be seen [48] that; $|\alpha_L(0)\rangle = |\alpha_R(0)\rangle = 0$.

- e) Eventually for any $r \in \mathbb{N}$ and to any given infinite dimensional vector $|\alpha\rangle$, we correspond an $r$-dimensional vector $|\alpha^{(r)}\rangle$ which its entries coincides with the first $r$ entries of $|\alpha\rangle$.

Now suppose that $m \geq 0$ is an integer and set $N = \frac{(m)^{m+1}-1}{m-1}(2m+1)^{2d}$. To define $\mathfrak{m}_N$ for translation-invariant product (5), one should use the Fourier modes over the torus $\mathbb{T}^{2d}$. For any $f \in \mathcal{S}(\mathbb{T}^{2d} \times \mathbb{R}^{2(n-d)})$ and for any two Fourier modes $\vec{p} = (p_1, \ldots, p_{2d}), \vec{q} = (p_1, \ldots, p_{2d}) \in \mathbb{Z}^{2d}$, set;

$$\mathfrak{m}_N(f)_{\vec{p}\vec{q}}^{(\mu)(\nu)}(x) := \frac{\alpha_R^{(m)}(ik_{\vec{p}})^{(\mu)}}{\sqrt{|\mu|!}} \frac{\alpha_L^{(m)}\left(i(k_{\vec{p}} - k_{\vec{q}})\right)^{(\nu)}}{\sqrt{|\nu|!}} \hat{f}(x, k_{\vec{p}} - k_{\vec{q}}),$$

(12)

with $k_{\vec{p}} = \frac{2\pi}{2\pi R}\vec{p} = \frac{1}{R}\vec{p}$, $x \in \mathbb{R}^{2(n-d)}$. Where $\hat{f}(x, k_{\vec{p}} - k_{\vec{q}})$ is the Fourier transformation of $f$ over $\mathbb{T}^{2d}$;

$$\hat{f}(x, k_{\vec{p}}) = \int_{y \in \mathbb{T}^{2d}} f(x, y) \frac{e^{-ik_{\vec{p}} \cdot y}}{(2\pi R)^{2d}}.$$

(13)

Note that the Haar measure on $\mathbb{T}^{2d}$ is chosen so that the volume of $\mathbb{T}^{2d}$ equals to unity. To keep the matrix of (12) finite dimensional, one has to restrict the Fourier modes of $\mathbb{T}^{2d}$; $\max(\{|p_i|\}_{i=1}^{2d} \cup \{|q_i|\}_{i=1}^{2d}) \leq m$. Also it is supposed that $|\mu|, |\nu| \leq m$. Therefore, $\mathfrak{m}_N(f)$ is an $N \times N$ matrix valued Schwartz function over $\mathbb{R}^{2(n-d)}$. Moreover, the limit of $R \to \infty$ is considered in the soul of definition (12) and thus $\mathfrak{m}_N(f)$ tends to a Hermitian matrix for real valued function $f \in \mathcal{S}(\mathbb{T}^{2d} \times \mathbb{R}^{2(n-d)})$. For simplicity from now on $\mathfrak{m}_N(f)$ is denoted by $[f]_N$.

As a definition $\star_N$ is defined by;

$$\int_{\mathbb{T}^{2d}} f_1 \star_N \ldots \star_N f_k := \lim_{R \to \infty} \xi_N tr\{[f_1]_N \ldots [f_k]_N\}$$

(14)

with the factor of $\xi_N = (\frac{2\pi R}{2m+1})^{2d}$, and for any $k \in \mathbb{N}$. In [48], it has been shown that (14) makes sense and thus $\star_N$ is defined definitely and independently from $k$. It can be easily seen that $\star_N$ tends to the translation-invariant product $\star$ defined by $\alpha$ as; $N(m) \to \infty$. Also by (14) it is obvious that;

$$\int_{\mathbb{T}^{2d}} f_1 \star_N \ldots \star_N f_{k-1} \star_N f_k = \int_{\mathbb{T}^{2d}} f_k \star_N f_1 \star_N \ldots \star_N f_{k-1}.$$

(15)

Moreover, (14) implies that [48];

$$f \star_N g = \pi \left( \sum_{p=0}^{m} \frac{1}{p!} \left( \sum_{i=1}^{m} \alpha_L(\vec{\partial})^i \otimes \alpha_R(\vec{\partial})^i \right)^p (f \otimes g) \right) + \cdots$$

(16)

for derivation operator $\vec{\partial} = (\partial_1, \ldots, \partial_{2d})$ and for the ordinary product map $\pi$. Note that the term of ... vanishes rapidly as; $N(m) \to \infty$. Actually $\star_N$ can be considered as a truncated form of $\star$ in order $m$. Indeed, (16) produces a generalized formula for $\theta$-expansion of Groenewold-Moyal product in matrix formulation. This expansion formula of translation-invariant noncommutative products provides a generalized version of Seiberg-Witten map [2] for translation-invariant noncommutative fields in the setting of matrix calculus.

Finally, it can also be checked that the Haar measure in (13) leads to;

$$\lim_{R \to \infty} (\frac{2\pi R}{2m+1})^{2d} tr[f]_N = \int_{\mathbb{T}^{2d}} f$$

(17)

for any smooth function $f$.

It should be noted that the matrix form of (12) is not the only possible formulation which leads to translation-invariant product (5), but it can be seen that (12) is the most compactified formula of matrix modeling for general translation-invariant noncommutative products over $C^\infty(M)$. For special translation-invariant products with finite dimensional $|\alpha_L\rangle$ (resp. $|\alpha_R\rangle$), (12) takes its simplest structure by equalizing $|\alpha^{(r)}\rangle$ and $|\alpha\rangle$ for any $r \in \mathbb{N}$. For example for the case of Groenewold-Moyal product, the matricial form (12) of $f \in \mathcal{S}(\mathbb{T}^{2d} \times \mathbb{R}^{2(n-d)})$ can be replaced by

$$[f]_{N\vec{p}\vec{q}}^{(\mu)(\nu)}(x) := \sum_{(\xi)(\zeta)} \frac{\sqrt{i\theta}^{(\mu)(\xi)} (ik_{\vec{p}})^{(\xi)}}{\sqrt{|\mu|!}} \frac{\left(i(k_{\vec{p}} - k_{\vec{q}})\right)^{(\zeta)} \sqrt{i\theta}^{(\zeta)(\nu)}}{\sqrt{|\nu|!}} \acute{f}(x, k_{\vec{p}} - k_{\vec{q}}),$$

(18)

for $\sqrt{i\theta}$ a square root of $i\theta$. Note that since $i\theta$ is a Hermitian matrix then there exists at least a solution for $\sqrt{i\theta}$.

The Lagrangian density of MMG theories for matter fields are defined in the setting of differential calculus on quantum groups. Actually the first approximation of $\mathcal{L}_{Matter}$ for MMG theories is defined by;

$$\mathcal{L}_{Matter} = i[\psi]_N^\dagger \gamma^0 \gamma^\mu [\partial_\mu \psi]_N - [\psi]_N^\dagger \gamma^0 \gamma^\mu T^a [A_\mu^a]_N [\psi]_N,$$

(19)

for spinor and gauge fields $\psi$ and $A$ respectively, modulo the Riemannian volume form of $M$ [48]. The other possible forms of the first approximation of matrix modeled Lagrangian density are;

$$\mathcal{L}_{Matter} = i[\psi]_N^\dagger \gamma^0 \gamma^\mu [\partial_\mu \psi]_N + [\psi]_N^\dagger \gamma^0 \gamma^\mu T^a [\psi]_N [A_\mu^a]_N,$$

(20)

and;

$$\mathcal{L}_{Matter} = i[\psi]_N^\dagger \gamma^0 \gamma^\mu [\partial_\mu \psi]_N - [\psi]_N^\dagger \gamma^0 \gamma^\mu T^a [[A_\mu^a]_N, [\psi]_N],$$

(21)

modulo the Riemannian volume form of $M$. On the other hand, the action of matter of MMG theories are given by

$$S_{Matter} := \int_{\mathbb{R}^{2(n-d)}} \xi_N tr(\mathcal{L}_{Matter}).$$

(22)

Therefore (22) in the limit of $N, R \to \infty$ leads to TNG theories on $M$. Indeed, matrix modeled Lagrangan densities (19)-(21), lead to TNG theories (7)-(9), respectively. This provides a semi-commutative setting to formulate TNG theories in the framework of matrix model formulations.

## 3. Consistent Anomalies and Schwinger Terms for Matrix Modeled Gauge Theories

As mentioned above the matrix modeled formulation of TNG theories lead to a matrix modeling generalization of $\theta$-expansion of Groenewold-Moyal noncommutative fields to any translation-invariant noncommutative star product $\star$. This consequently results in matricial generalization of Seiberg-Witten map for any translation-invariant noncommutative field. The truncated form of star product $\star$, $\star_N$, defined by (16) in matrix model formulation, leads to higher derivations of matter and gauge fields to appear in the matrix modeled Lagrangian densities (19)-(21). Obviously this results in worse ultraviolet and consequently nonrenormalizability behaviors of MMG theories similar to $\theta$-expanded versions of Groenewold-Moyal noncomutative gauge theories [26]. Therefore, MMG theories are nonrenormalizable while the renormalizability of TNG theories is not clear yet [5].

Although, the relation of the renormalizability of TNG and MMG theories is not clear yet, but here it is assumed that the anomalous behaviors of a TNG theory can be considered as the limit of those of a MMG theory [48]. According to [49], modified partition functions enable one to calculate all the anomalous behaviors of an ordinary gauge theory derived by descent equations via path integral methods. In [48], it has already been shown that this is also true for MMG theories. Consequently, modified partition functions of MMG theories produce an algebraic setting to extract the consistent anomalous terms of TNG theories via the formalism of path integral quantization.

According to [49] to extract the consistent anomalous terms from modified partition functions, the infinitesimal gauge transformations are used in the formalism of BRST derivations. Therefore, following [48], the anomalous terms are the solutions of a matricial form or a noncommutative version of descent equations initiated by the matricial counterpart of Chern character appearing in modified partition functions of MMG theories. Thus consistent anomalies and consistent Schwinger terms of MMG theories due to path integral quantization of modified partition functions are in complete agreement with those derived by noncommutative version of descent equations in the setting of noncommutative geometry [48, 51] (see also [52, 53]).

The infinitesimal matrix modeled gauge transformations $[\alpha]_N \otimes 1$ and $1 \otimes [\alpha]_N$, $\alpha \in \mathcal{S}(M)$, in the setting of matrix model formulation [48] of Abelian gauge theory, lead to two apparently independent currents due to Noether's theorem, $[\psi_j]_N [\psi_i]_N^\dagger (\gamma^0 \gamma^\mu)_{ij}$ and $[\psi]_N^\dagger \gamma^0 \gamma^\mu [\psi]_N$. Thus, it seems that there are two linearly independent anomalies for each of these currents. This fact is intimately related to the expressed paradox in [44] for anomalies of axial invariant and covariant currents in NCQED. This relation can be illustrated more apparently by using the anomalous behaviors of the first approximation of Abelian $N \times N$-MMG theory, calculated via the idea of modified partition functions in the consistent formalism. These results in the limit of $N \to \infty$ are shown to be equal to those of invariant and covariant currents in Groenewold-Moyal NCQED [36]. Thus the abnormal structures of anomalies of invariant and covariant currents are discussed in the setting of matrix model formulations. Moreover, the consistent Schwinger term for invariant/covariant currents in NCQED is extracted similarly in the matrix model formulation and it is also shown that it is in absolute agreement with the results of Feynman diagram calculations [45].

For simplicity, to extract the consistent anomalies for axial invariant and covariant currents, we restrict ourselves to vector invariant and covariant currents and follow [50] to derive the results for the axial ones.

Consider the following action;

$$S_{Matter}(A) = \int_{\mathbb{R}^{2(n-d)}} \xi_N \, tr\{[\overline{\psi}]_N i\gamma^\mu [\partial_\mu \psi]_N - [\overline{\psi}]_N \gamma^\mu [A_\mu]_N [\psi]_N\}$$

(23)

with $A = [A_\mu]_N \otimes 1 \; dx^\mu$, for real valued Schwartz functions $A_\mu$. As it was stated above, the action (23) can be rewritten by;

$$S_{Matter}(A) = \int_{\mathbb{R}^{2(n-d)} \times \mathbb{T}^{2d}} \bar{\psi}_\alpha \star_N i\partial_\mu \psi_\beta \gamma^\mu_{\alpha\beta} - \bar{\psi}_\alpha \star_N A_\mu \star_N \psi_\beta \gamma^\mu_{\alpha\beta} , \tag{24}$$

which yields the familiar Lagrangian;

$$\mathcal{L}_{Matter}(\psi) = \bar{\psi}_\alpha \star_N i\partial_\mu \psi_\beta \gamma^\mu_{\alpha\beta} - \bar{\psi}_\alpha \star_N A_\mu \star_N \psi_\beta \gamma^\mu_{\alpha\beta} , \tag{25}$$

over $M$. Now consider the gauge transformation $e^{ti\tilde{\alpha}}$ with $\tilde{\alpha} = [\alpha]_N \otimes 1$ for $\alpha$ a real Schwartz function over $M$ and $t \in \mathbb{R}$. Gauge transformation defines a right action of $e^{ti\tilde{\alpha}}$ on gauge field $A$ by;

$$A \triangleleft e^{ti\tilde{\alpha}} := ie^{-ti\tilde{\alpha}} de^{ti\tilde{\alpha}} + \mathrm{Ad}_{e^{-ti\tilde{\alpha}}}(A)$$
$$= ie^{-ti\tilde{\alpha}} de^{ti\tilde{\alpha}} + e^{-ti\tilde{\alpha}} A e^{ti\tilde{\alpha}} . \tag{26}$$

The derivation of $\frac{d}{dt}\big|_{t=0} S_{Matter}(A \triangleleft e^{ti\tilde{\alpha}}) := \int_{\mathbb{R}^{2(n-d)} \times \mathbb{T}^{2d}} \delta_{i\tilde{\alpha}} \mathcal{L}_{Matter}$, leads to;

$$\delta S_{Matter}(A) = -\int_{\mathbb{R}^{2(n-d)}} \xi_N \; tr\{\overline{[\psi]}_N \gamma^\mu [\partial_\mu \alpha]_N [\psi]_N - i\overline{[\psi]}_N \gamma^\mu [[\alpha]_N, [A_\mu]_N][\psi]_N\} . \tag{27}$$

A direct calculation shows that;

$$\delta S_{Matter}(A) = -\int_M \alpha \star_N (\partial_\mu(\psi_b \star_N \bar{\psi}_a \gamma^\mu_{ab}) - i[\psi_b \star_N \bar{\psi}_a \gamma^\mu_{ab}, A_\mu]_{\star_N}) , \tag{28}$$

where the minus sign on the right hand side of (28) comes from the commutation of Grassmannian fields $\psi$ and $\bar{\psi}$ in the path integral setting. The classical equations of motion show that; $\delta S_{Matter}(A) = 0$. Specially, setting $J^\mu = \psi_b \star_N \bar{\psi}_a \gamma^\mu_{ab}$, replacing $\star$ by its $\alpha$-cohomologous harmonic translation-invariant product [48] and taking the functional derivative with respect to $\alpha(x, y)$, $(x, y) \in \mathbb{T}^{2d} \times \mathbb{R}^{2(n-d)}$, the classical conservation law for covariant current is derived;

$$D(J)(x,y) = \partial_\mu J^\mu(x,y) - i[J^\mu(x,y), A_\mu(x,y)]_{\star_N} = 0 .$$

(29)

Note that, the elegant property of harmonic translation-invariant products in integration, i.e. $\int f \star g = \int fg$ and the cohomological spirit of quantum corrections in TNG theories [48], plays the crucial role in extracting the classical conservation law (29), consistently.

To find the consistent anomaly for covariant current we follow [48] (see also [49, 51] and the references therein). The modified partition function of matrix modeled Abelian gauge theory is [48];

$$Z_M(A) = Z(A) e^{-i \int_{\mathbb{R}^{2(n-d)+1}} \eta_{n,d} \Omega_{2n+1}(A)} ,$$

(30)

for $d\Omega_{2n+1}(A) = \Omega_{2n+2}(A)$ with $\Omega_{2n+2}(A) = Tr\{\hat{R}^{n+1}\}$ the $(n+1)$th Chern character for the curvature

$$\hat{R} = \frac{1}{2}\left(i[\partial_{[\mu} A_{\nu]}]_N \otimes 1_{N\times N} - ([A_{[\mu}]_N \otimes [A_{\nu]}]_N) \otimes 1_{N\times N}\right) dx^\mu \wedge dx^\nu$$

(31)

and for $\eta_{n,d} = \frac{c_n}{(2\pi R)^{2d}}$ a topological factor proportional to $c_n$ (see [49] and the references therein). Moreover, $Tr$ is a trace operator given by;

$$Tr\{\varpi\} := \sum_{i,\alpha,\beta} \xi_N tr\{\prod_j a^i_{\alpha_j}\} \xi_N tr\{\prod_{\downarrow j}(-b^i_{\beta_j}))\} ,$$

(32)

for $\varpi = \sum_i \varpi^i_a \otimes \varpi^i_b$ with $\varpi^i_x = \sum_\alpha x^i_{\alpha_1} \otimes ... \otimes x^i_{\alpha_{n_\alpha}}$, $x^i_{\alpha_j} \in \mathfrak{m}_N(\mathcal{S}(M)) \cup \mathbb{C}1_{N\times N}$ $x = a, b$. But $\prod_{i=1}^k c_i$ is the ordinary notation for $c_1 ... c_k$ while $\prod_{\downarrow i=1}^k c_i$ is used for $c_k ... c_1$.

Gauge invariance of $Z_M(A)$ implies that at the quantum level one finds;

$$\delta S_{Matter}(A) = \int_{\mathbb{R}^{2(n-d)}} \eta^N_{n,d} Tr\{\Omega^1_{2n}(A)(i\tilde{\alpha})\} ,$$

(33)

where $\Omega^1_{2n}(A)$ is the consistent anomaly derived by descent equations; $d\Omega_{2n+1}(A) = Tr\{\hat{R}^{n+1}\}$ and $d\Omega^1_{2n}(A) = \delta\Omega_{2n+1}(A)$, for $\delta$ the matricial form of noncommutative BRST derivative operator defined by [48];

$$\delta A = d\omega - iA\omega - i\omega A \;, \quad \delta\omega = -i\omega^2 \;.$$

(34)

The crucial point is that the descent equations holds for deRham-BRST forms over $\mathbb{R}^2 \times M$ up to the $(2d+1)$th term $\Omega^{2d}_{2(n-d)+1}(A)$. Actually, since $H^{2n-k}_{deR}(\mathbb{R}^2 \times M) = 0$ for $k < 2(n-d)$, any closed $2n-k$-form over $\mathbb{R}^2 \times M$ is exact if $k < 2(n-d)$, while $H^k_{deR}(\mathbb{R}^2 \times M) = \mathbb{R}^{\binom{2d}{k}}$ for $k \leq 2d$, hinders the exactness of closed forms generally. Thus according to [49], (28) and (33) lead to;

$$\int_M \alpha \star_N (\partial_\mu(\psi_b \star_N \bar{\psi}_a \gamma^\mu_{ab}) - i[\psi_b \star_N \bar{\psi}_a \gamma^\mu_{ab}, A_\mu]_{\star_N})$$

$$= - \int_{\mathbb{R}^{2(n-d)}} \eta_{n,d} \, Tr\{\Omega^1_{2n}(A)(i\tilde{\alpha})\}$$

(35)

Indeed for a 4-dimensional $N \times N$-matrix modeled QED over $M = \mathbb{R}^2 \times \mathbb{T}^2$ one finds;

$$\int_{\mathbb{R}^2} \frac{1}{(2\pi R)^2} Tr\{\Omega^1_4(A)(i\tilde{\alpha})\}$$

$$= \int_{\mathbb{R}^2 \times \mathbb{T}^2} \epsilon^{\mu\nu\sigma\lambda} \; \alpha \star_N (i \, \partial_\mu A_\nu \star_N \partial_\sigma A_\lambda - \frac{1}{2} \partial_\mu(A_\nu \star_N A_\sigma \star_N A_\lambda))$$

(36)

Thus, since the quantum corrections in a TNG theory is entirely related to the $\alpha$-cohomology class of its translation-invariant star product, then $\star$ can be replaced by its $\alpha$-cohomologous harmonic translation-invariant product. Therefore, (28) and (36) can be revised to give;

$$\int_{\mathbb{R}^2 \times \mathbb{T}^2} \alpha \, (\partial_\mu (\psi_b \star_N \bar{\psi}_a \gamma^\mu_{ab}) - i[\psi_b \star_N \bar{\psi}_a \gamma^\mu_{ab}, A_\mu]_{\star_N})$$

$$= \frac{1}{24\pi^2} \int_{\mathbb{R}^2 \times \mathbb{T}^2} \epsilon^{\mu\nu\sigma\lambda} \, \alpha \, (i \, \partial_\mu A_\nu \star_N \partial_\sigma A_\lambda - \frac{1}{2} \partial_\mu (A_\nu \star_N A_\sigma \star_N A_\lambda)).$$

(37)

Taking the functional derivative of both sides of (37) with respect to $\alpha(x, y)$ leads to;

$$D\langle J(x,y) \rangle = \partial_\mu \langle J^\mu(x,y) \rangle + i[\langle J^\mu(x,y) \rangle, A_\mu(x,y)]_{\star_N}$$

$$= \frac{1}{24\pi^2} \epsilon^{\mu\nu\sigma\lambda} \left( i\partial_\mu A_\nu \star_N \partial_\sigma A_\lambda - \frac{1}{2} \partial_\mu (A_\nu \star_N A_\sigma \star_N A_\lambda) \right)(x,y).$$

(38)

Since NCQED with star product $\star$ is the limit theory of the action (23), then it can be naturally assumed that (38) gives hand the consistent anomaly of covariant current as $N, R \to \infty$ which lead to $\star_N \to \star$. It can be seen that according to [54] adding a set of compatible renormalizing counter terms to the effective action will yield the covariant current to be anomaly free, while this simultaneously causes the axial covariant current to bear the whole anomaly contribution. Thus following [54] one fins that;

$$D\langle J^5(x,y) \rangle = \frac{1}{16\pi^2} \epsilon^{\mu\nu\sigma\lambda} F_{\mu\nu} \star_N F_{\sigma\lambda}(x,y) \quad , \quad D\langle J(x,y) \rangle = 0 \, ,$$

(39)

in complete agreement with (3).

Now consider the gauge transformation $e^{ti\tilde{\beta}}$, $t \in \mathbb{R}$, with $\tilde{\beta} = 1 \otimes [\beta]_N$ for $\beta$ a real Schwartz function over $M$ and calculate $\frac{d}{dt}\big|_{t=0} Z_M(A \triangleleft e^{ti\tilde{\beta}})$. Then, since $Z_M(A)$ is gauge invariant one finds that;

$$\int_{\mathbb{R}^{2(n-d)}} \xi_N tr\{\overline{[\psi]}_N \gamma^\mu [\psi]_N [\partial_\mu \beta]_N - i[[\beta]_N, \overline{[\psi]}_N \gamma^\mu [A_\mu]_N [\psi]_N]\}$$

$$= \int_{\mathbb{R}^{2(n-d)}} \eta_{n,d} \, Tr\{\Omega^1_{2n}(A)(i\tilde{\beta})\}.$$

(40)

A direct calculation shows that (40) leads to;

$$\int_M \beta \star_N (\partial_\mu(\bar\psi_a \star_N \psi_b \gamma^\mu_{ab})) = \int_{\mathbb{R}^{2(n-d)}} \eta_{n,d}\, Tr\{\Omega^1_{2n}(A)\,(i\tilde\beta)\}.$$

(41)

Similarly considering the cohomologic property of quantum corrections of TNG theories, one can replace $\star_N$ (resp. $\star$) in (41) by its $\alpha$-cohomologous harmonic translation-invariant star product. Setting $j^\mu = \bar\psi_a \star_N \psi_b \gamma^\mu_{ab}$ and taking the functional derivative $\frac{\delta}{\delta\beta(x,y)}$, of the left hand side of (41), the classical conservation law for covariant current is derived;

$$\partial_\mu j^\mu(x,y) = 0.$$

(42)

On the other hand, in the case of 4-dimensional space-time ($M = \mathbb{R}^2 \times \mathbb{T}^2$) the right hand side of (41) leads to;

$$\int_{\mathbb{R}^2} \frac{1}{(2\pi R)^{2d}} Tr\{\Omega^1_4(A)(i\tilde\beta)\}$$

$$= -\int_{\mathbb{R}^2} \xi_N tr\{d(i[A]_N d[A]_N - \frac{1}{2}[A]_N[A]_N[A]_N)\} \times \frac{1}{(2\pi R)^{2d}} \xi_N tr[\beta]_N.$$

(43)

Functional derivation of (41) and (43) with respect to $\beta(x,y)$ gives hand the anomaly of invariant current;

$$\partial_\mu \langle j^\mu(x,y) \rangle$$

$$= -\frac{1}{24\pi^2 (2\pi R)^2} \int_{z \in \mathbb{T}^{2d}} \epsilon^{\mu\nu\sigma\lambda} \left( i\partial_\mu A_\nu \star_N \partial_\sigma A_\lambda - \frac{1}{2}\partial_\mu(A_\nu \star_N A_\sigma \star_N A_\lambda) \right)(x,z).$$

(44)

Therefore, the consistent anomaly of invariant current vanishes as $R \to \infty$ because of the factor $\frac{1}{(2\pi R)^2}$. Thus the invariant current seems to be anomaly free [39]. On the other hand, the right hand side of (44) is independent of $y$ and thus $\partial_\mu \langle j^\mu(x,y) \rangle$ defines a constant function on $\mathbb{T}^2$. Thus $\partial_\mu \langle j^\mu(x,y) \rangle$ contributes only to the zero mode of the Fourier expansion, the result of which had been discovered in [39, 45]. Actually the Fourier transformation of $\partial_\mu \langle j^\mu(x,y)\rangle$ lives over a Lebesgue measure null set of the phase space of $M = \mathbb{R}^4$, $(R \to \infty)$, the fact of which had been observed before [39]. It can obviously be seen that (44) solves the paradox of which was pointed out in [44]. The authors of [44] found out that if the invariant current is anomaly free, then the following natural equation cannot be hold;

$$\int_{y \in \mathbb{T}^{2d}} D(J)(x,y) = -\int_{y \in \mathbb{T}^{2d}} \partial_\mu j^\mu(x,y)(x,y).$$

(45)

It can be seen that (38) and (44) naturally solve the denoted paradox of [44] for the Groenewold-Moyal NCQED, as the limit theory of action (23). It can be seen that in addition to Groenewold-Moyal NCQED, (45) holds for any translation-invariant NCQED.

One can also calculate the consistent Schwinger term of covariant and invariant currents in NCQED equipped with a general translation-invariant product $\star$. As it was shown above such terms should be initially calculated for MMG theory of (23) and then be evaluated in the limit of $N, R \to \infty$. Following the manipulations of [49] one finds that;

$$\langle [J^0(t,x_1,y_1), j^0(t,x_2,y_2)]\rangle|_{anomalous}$$
$$= i\eta_{n,d} \frac{\delta}{\delta_{2(n-d)}\alpha(x_1,y_1)} \frac{\delta}{\delta_{2(n-d)}\beta(x_2,y_2)} \int_{x \in \mathbb{R}^{2(n-d)-1}} Tr\{(\Omega^2_{2n-1}(A)(i\tilde{\alpha}, i\tilde{\beta}))\}(t,x),$$

(46)

for $(x_i, y_i) \in \mathbb{R}^{2(n-d)-1} \times \mathbb{T}^{2d}$, $i = 1,2$. Note that the infinitesimal gauge transformations $\tilde{\alpha} = [\alpha]_N \otimes [1]_N$ and $\tilde{\beta} = [1]_N \otimes [\beta]_N$ are similarly considered as above.

Similarly, considering the cohomologic structure of quantum corrections of TNG theories, one can naturally replace the star product $\star$ by its $\alpha$-cohomologous harmonic translation-invariant product. Therefore, for the case of $M = \mathbb{R}^2 \times \mathbb{T}^2$, (46) takes the form of

$$\langle [J^0(t,x_1,y_1), j^0(t,x_2,y_2)]\rangle|_{anomalous}$$
$$= \frac{1}{24\pi^2(2\pi R)^2} \int_{x \in \mathbb{R}} \{\int_{y \in \mathbb{T}^2} \epsilon^{ijk} \partial_i \delta^{(3)}_{x,y}(x_1,y_1) \partial_j A_k(x,y) \times \int_{z \in \mathbb{T}^2} \delta^{(3)}_{x,z}(x_2,y_2)\},$$

(47)

for $i,j,k = 1,2,3$, and consequently we have;

$$\langle [J^0(t,x_1,y_1), j^0(t,x_2,y_2)] \rangle |_{anomalous}$$

$$= \frac{\epsilon^{1jk}}{24\pi^2 (2\pi R)^2} (\delta'(x_2 - x_1) - \delta(x_2 - x_1)\partial_1)\partial_j A_k(x_2, y_1),$$

(48)

for $y_1, y_2 \in \mathbb{T}^2$ and $x_1, x_2 \in \mathbb{R}$. As it is seen from (48), the explicit form of consistent Schwinger term is independent of $N$ and proportional to $\frac{1}{R^2}$, in a complete agreement with the results of [45]. Obviously, taking the limit of $N, R \to \infty$, kills the calculated consistent Schwinger term for NCQED. Although (48) shows that the consistent Schwinger term of commutation of invariant and covariant currents, vanishes for any translation-invariant NCQED over $\mathbb{R}^4$ ($R \to \infty$), but the integration of the consistent Schwinger term over $y_2 \in \mathbb{T}^2$ survives similarly.

# Conclusion

In this article the anomalous behaviors of invariant and covariant currents of general translation-invariant NCQED have been exhaustively discussed in the setting of matrix model formulation. It was shown that the anomalies of invariant and covariant currents of the first approximation of Abelian MMG theory are in complete agreement with those of Groenewold-Moyal NCQED extracted using loop calculations. Actually it is shown that in the formulation of MMG theories the invariant and covariant currents are two different features of a single Noether current due to noncommutative gauge symmetry. This naturally illustrates the paradox of [44] in different anomalous behaviors of invariant and covariant currents in Groenewold-Moyal NCQED for any general translation-invariant star product.

Moreover, the consistent Schwinger term of commutator of invariant and covariant currents was calculated for any NCQED with general translation-invariant star product. It was shown that the result is in complete agreement with that derived from loop calculations in Groenewold-Moyal NCQED.

To derive all the results above, we used the elegant theorem of equality of quantum corrections of two NCQED theories defined by two $\alpha$-cohomolog-ous translation-invariant star product. We used this theorem to replace the translation-invariant star product $\star$ with its $\alpha$-cohomologous harmonic counterpart.

## Acknowledgement

The author acknowledges his gratitude to F. Ardalan for discussions and comments. Moreover, I should say my gratitude to E. Langmann for his beneficial hints and discussions. On the other hand, I should confess that this article owes most of its appearance to S. Ziaee for many things. Finally my special thanks go to A. Shafiei Deh Abad for his beneficial hints and drastic ideas.